\RequirePackage{ifpdf}
\ifpdf % We are running pdfTeX in pdf mode
\documentclass[pdftex]{sigma}
\else
\documentclass{sigma}
\fi

\def\CC{\mathbb{C}}
\def\RR{\mathbb{R}}
\def\ZZ{\mathbb{Z}}
\def\SS{\mathbb{S}}
\def\NN{\mathbb{N}}

\def\res{{\mathop{\rm res}}}

\def\D{\displaystyle}

\begin{document}
\allowdisplaybreaks

\renewcommand{\PaperNumber}{001}

\FirstPageHeading

\ShortArticleName{Lax Integrable Supersymmetric Hierarchies on
Extented Phase Spaces}

\ArticleName{Lax Integrable Supersymmetric Hierarchies \\ on
Extented Phase Spaces}

\Author{Oksana~Ye.~HENTOSH} \AuthorNameForHeading{O.Ye.~Hentosh}

\Address{Institute for Applied Problems of Mechanics and
Mathematics,\\ National Academy of Sciences of Ukraine,
3B Naukova Str., Lviv, 79060 Ukraine} % Address of First Author
\Email{\href{mailto:dept25@iapmm.lviv.ua}{dept25@iapmm.lviv.ua}}

\ArticleDates{Received October 27, 2005, in final form December
21, 2005; Published online January 04, 2006}

\Abstract{We obtain via B\"acklund transformation the Hamiltonian
representation for a Lax type nonlinear dynamical system hierarchy
on a dual space to the Lie algebra of
super-integral-diffe\-rential operators of one anti\-commuting
variable, extended by evolutions of the corresponding spectral
problem eigenfunctions and adjoint eigenfunctions, as well as for
the hierarchies of their additional symmetries. The relation of
these hierarchies with the integrable by Lax
($2|1+1$)-dimen\-sional supersymmetric Davey--Stewartson system is
investigated.}


\Classification{35Q53; 35Q58; 37K10; 37K30; 37K35; 58A50}

\section{Introduction}

Since the paper of M.~Adler~\cite{hentosh1} there was an
understanding that Lax forms for a wide class of integrable
nonlinear dynamical system hierarchies on functional
manifolds~\cite{hentosh2,hentosh3,hentosh4,hentosh5} and their
supersymmetric analogs~\cite{hentosh6,hentosh7} could be
considered as Hamiltonian flows on dual spaces to the Lie algebra
of integro-differential operators. Those flows are generated by
the ${\cal R}$-deformed canonical Lie--Poisson bracket and Casimir
functionals as Hamiltonian functions
(see~\cite{hentosh1,hentosh8,hentosh9}). For a concrete
integro-differential operator every Hamiltonian flow of such a
type can be written as a compatibility condition for the
corresponding isospectral problem in the case of an arbitrary
eigenfunction and the suitable evolution of this function. Thus,
the existence problem of a Hamiltonian representation for the Lax
type hierarchy, extended by the evolutions of a finite set of
eigenfunctions and appropriate adjoint eigenfunctions, arises.
In~\cite{hentosh10,hentosh11,hentosh12} it was solved for the Lie
algebra of integro-differential operators by use of the Casimir
functionals' invariant property under some Lie--B\"acklund
transformation. Analogously we obtain in this paper the
Hamiltonian representation of the extended Lax type system
hierarchy for the Lie algebra of super-integro-differential
operators of one anticommuting variable.

The hierarchies of additional or ``ghost''
symmetries~\cite{hentosh12a} for the extended Lax type system are
also proved to be Hamiltonian. It is established that every
additional symmetry hierarchy is gene\-rated by the tensor product
of the ${\cal R}$-deformed canonical Lie--Poisson bracket with the
Poisson bracket on a finite-dimensional superspace, possessing an
odd supersymplectic structure~\cite{hentosh14,hentosh15}, and all
natural powers of one eigenvalue from the mentioned above finite
set as Hamiltonian functions. The additional symmetry hierarchy is
used for introducing one more commuting va\-riab\-le into
$(1|1+1)$-dimensional supersymmetric nonlinear dynamical systems
with preserving their integrability by Lax. By means of this
approach a $(2|1+1)$-dimensional supersymmetric analog of the
Davey--Stewartson system~\cite{hentosh5,hentosh16,hentosh17} and
its triple linearization of a Lax type are found.

\section{The general algebraic scheme}

Let ${\cal G}$ be a Lie algebra of scalar
super-integral-differential operators~\cite{hentosh6} of one
anticommuting variable $\theta$ ($\theta^2=0$):
\[
a:=\partial^m+\sum_{j<2m}a_jD_\theta^j, \qquad m\in\NN,
\]
where the symbol $\partial:=\partial/\partial x$ designates
differentiation with respect to the independent variable $x\in \RR
/2\pi \ZZ\simeq \SS^1$, $a_j:=a_j(x,\theta)=a_j^0(x)+\theta
a_j^1(x)$, $j\in \ZZ$, are smooth superfield functions
(superfunctions), and the superderivative
$D_\theta:=\partial/\partial \theta +\theta \partial/\partial x$,
for which $D_\theta^2=\partial$, satisfies the following
 relation for any smooth superfield functions $u$ and~$v$:
\[
D_\theta (uv)=(D_ \theta u)v+(-1)^{p(u)}u(D_\theta v),
\]
where $p(u)$ is a parity of an arbitrary superfunction $u$, which
is equal to $0$ for $u$, being even, and one for $u$, being odd.

The usual Lie commutator on $ {\cal G}$ is defined as
\[
[a, b]:=a\circ b-b\circ a
\]
for all $a, b\in {\cal G}$, where ``$\circ$'' is an associative
product of super-integro-differential operators. On the Lie
algebra $ {\cal G}$ there exists the $ad$-invariant nondegerated
symmetric bilinear form:
\begin{gather}\label{eq1}
( a, b):=\int_{0}^{2\pi}\res _{D_\theta}\, ( a\circ b)\ dx,
\end{gather}
where $\res _{D_\theta}$-operation for all $ a\in {\cal G}$ is
given by the expression:
\[
\res _{D_\theta}\, a := a_{-1}.
\]
By means of the scalar product~\eqref{eq1} the Lie algebra $ {\cal
G}$ is transformed into a metrizable one. As a consequence, its
dual linear space of scalar super-integro-differential operators
${\cal G}^*$ is identified with the Lie algebra ${\cal G}$, that
is ${\cal G}^*\simeq {\cal G}$.

The linear subspaces $ {\cal G}_+\subset {\cal G}$ and ${\cal G}_-
\subset  {\cal G}$
\begin{gather}
{\cal G}_+:= \left\{ a:=\partial^m+\sum_{j=0}^{2m-1}a_j
D_{\theta}^j:\ j=\overline{0,2m-1}\right\},
\nonumber \\
{\cal G}_-:= \left\{ b:=\sum_{l>0}^{\infty}b_l D_{\theta}^{-l}:\
l\in \NN \right\},\label{eq2}
\end{gather}
where $a_j$ and $b_l$ are smooth superfunctions, forms Lie
subalgebras in ${\cal G}$ and ${\cal G} = {\cal G}_+\oplus {\cal
G}_-$. Because of the splitting of ${\cal G}$ into the direct
sum~\eqref{eq2} of its Lie subalgebras one can construct
a~Lie--Poisson structure~\cite{hentosh1,hentosh8,hentosh9} on
${\cal G}^*$, using the special linear endomorphism ${\cal R}$ of
${\cal G}$:
\[
{\cal R} :=(P_+-P_-)/2, \qquad P_{\pm} {\cal G} := {\cal G}_{\pm},
\qquad P_{\pm} {\cal G}_{\mp}=0.
\]
For any smooth by Frechet functionals $\gamma,\mu\in {\cal D}(
{\cal G}^*)$ the Lie--Poisson bracket on ${\cal G}^*$ is given by
the expression:
\begin{gather}\label{eq3}
\left\{ \gamma,\mu \right\}_{{\cal R}}(l)= \left ( l ,[\nabla
\gamma(l),\nabla \mu(l)]_{{\cal R}}  \right ),
\end{gather}
where $ l\in {\cal G}^*$ and for all $ a,\ b\in {\cal G}$ the
${\cal R}$-deformed commutator has the form:
\begin{gather}\label{eq4}
[ a, b]_{{\cal R}}:=[{\cal R} a, b]+[ a,{\cal R} b].
\end{gather}
The linear space ${\cal G}$ with the commutator~\eqref{eq4} also
becomes a Lie algebra. The gradient $\nabla\gamma (l)\in {\cal G}$
of some functional $\gamma\in {\cal D} ({\cal G}^*)$ at the point
$ l\in {\cal G}^*$ with respect to the scalar product~\eqref{eq1}
is defined as
\[
\delta\gamma (l):= \left (\nabla \gamma (l),\delta l\right ),
\]
where the linear space isomorphism ${\cal G}\simeq {\cal G}^*$ is
taken into account.

Every Casimir functional $\gamma\in I (G^*)$, being invariant with
respect to ${\rm Ad}^*$-action of the corresponding Lie group $G$,
obeys the following condition at the point $l\in {\cal G}^*$:
\begin{gather}\label{eq5}
[l , \nabla\gamma (l)]=0.
\end{gather}
The relationship~\eqref{eq5} is satisfied by the hierarchy of
functionals $\gamma_n\in I(G^*)$, $n\in \ZZ_+$, taking the forms:
\begin{gather}\label{eq6}
\gamma_n (l)= \frac 1{n+1} (l^{1/m},l^{n/m}).
\end{gather}

The Lie--Poisson bracket~\eqref{eq3} generates the hierarchy of
Hamiltonian dynamical systems on~${\cal G}^*$:
\begin{gather}\label{eq7}
d l/dt_n:= [{\cal R}\nabla\gamma_n (l), l]= [(\nabla\gamma_n
(l))_+, l],
\end{gather}
with the Casimir functionals~\eqref{eq6} as Hamiltonian functions.

The latter equation is equivalent to the usual commutator Lax type
representation. It is easy to verify that for every $n\in \ZZ_+$
the relationship~\eqref{eq7} is a compatibility condition for such
linear integral-differential equations:
\begin{gather}\label{eq8}
l f= \lambda f,
\end{gather}
and
\begin{gather}\label{eq9}
df/dt_n= (\nabla\gamma_n (l))_+f,
\end{gather}
where $\lambda\in\CC$ is a spectral parameter, $f\in W^{1|0}:=
L_\infty (\SS^1\times \Lambda_1;\CC^{1|0})$ if $f$ is an even
superfunction and $f\in W^{0|1}:= L_\infty (\SS^1\times
\Lambda_1;\CC^{0|1})$ if $f$ is an odd one. Here
$\Lambda:=\Lambda_0\oplus \Lambda_1$ is a Grassmann algebra over
$\CC$, $\Lambda_0\supset \RR$. The associated with \eqref{eq9}
dynamical system for the adjoint superfunction $f^*$ takes the
form:
\begin{gather}\label{eq10}
df^*/dt_n=-(\nabla\gamma_n (l))^*_+ f^*,
\end{gather}
where $(f,f^*)^T\in W^{1|1}:= L_\infty (\SS^1\times
\Lambda_1;\CC^{1|1})$ or $(f^*,f)^T\in W^{1|1}$ and superfunction
$f^*$ is a solution of the adjoint spectral problem:
\[
l^* f^*=\lambda f^*.
\]
The objects of further investigations are some algebraic
properties of equation~\eqref{eq7} together with $2N\in \NN$
copies of equation~\eqref{eq9}:
\begin{gather}
df_i/dt_n= (\nabla\gamma_n (l))_+f_i, \nonumber\\
d\Phi_i/dt_n= (\nabla\gamma_n (l))_+\Phi_i,\label{eq11}
\end{gather}
for even $f_i\in W^{1|0}$ and odd $\Phi_i\in W^{0|1}$
eigenfunctions of the spectral problem~\eqref{eq8}, corresponding
to the eigenvalues $\lambda_i$, $i=\overline{1,N}$, and the same
number of copies of equation~\eqref{eq10}:
\begin{gather}
df_i^*/dt_n=-(\nabla\gamma_n (l))^*_+ f_i^*, \nonumber \\
d\Phi_i^*/dt_n=-(\nabla\gamma_n (l))^*_+ \Phi_i^*,\label{eq12}
\end{gather}
for corresponding odd $f_i^*\in W^{0|1}$ and even $\Phi_i^*\in
W^{1|0}$ adjoint eigenfunctions, as a coupled evolution system on
the space ${\cal G}^*\oplus W^{2N|2N}$.

\section[Tensor product of Poisson structures and its B\"acklund transformation]{Tensor product
of Poisson structures\\ and its B\"acklund transformation}

To compactify the description below one shall use the following
designation of the left gradient vector:
\[
\nabla\gamma (\tilde l, \tilde f_i, \tilde \Phi_i^*, \tilde f_i^*,
\tilde \Phi_i):= \left ( \frac{\delta\gamma}{\delta\tilde l},\,
\frac{\delta\gamma}{\delta\tilde f_i},\,
\frac{\delta\gamma}{\delta\tilde \Phi_i^*},\,
\frac{\delta\gamma}{\delta\tilde f_i^*},\,
\frac{\delta\gamma}{\delta\tilde \Phi_i}\right )^T,
\]
where $i=\overline{1,N}$, at a point $(\tilde l, \tilde f_i,
\tilde \Phi_i^*, \tilde f_i^*, \tilde \Phi_i)^T\in {\cal
G}^*\oplus W^{2N|2N}$ for any smooth functional $\gamma\in{\cal D}
({\cal G}^*\oplus W^{2N|2N})$.

On the spaces $ {\cal G}^*$ and $W^N\oplus W^{*N}$ there exist a
Lie--Poisson structure~\cite{hentosh1,hentosh8,hentosh9}
\begin{gather}\label{eq13}
\delta\gamma/\delta\tilde l:\stackrel{\tilde\Theta}{\to} \left
[\tilde l ,\left (\frac{\delta\gamma}{\delta\tilde l}\right
)_+\right ] - \left [\tilde l ,\frac{\delta\gamma}{\delta\tilde
l}\right ]_+,
\end{gather}
where $\tilde\Theta:{\cal G}\to {\cal G}^*$, at a point $\tilde
l\in {\cal G}^*$ and the canonical Poisson
structure~\cite{hentosh14,hentosh15}
\begin{gather}\label{eq14}
\left ( \frac{\delta\gamma}{\delta\tilde f_i},\,
\frac{\delta\gamma}{\delta\tilde \Phi_i^*},\,
 \frac{\delta\gamma}{\delta\tilde f_i^*},\, \frac{\delta\gamma}{\delta\tilde \Phi_i}\right )^T:
\stackrel{\tilde J}{\to} \left ( -\frac{\delta\gamma}{\delta\tilde
f_i^*},\, \frac{\delta\gamma}{\delta\tilde \Phi_i},\,
\frac{\delta\gamma}{\delta\tilde f_i},\,
-\frac{\delta\gamma}{\delta\tilde \Phi_i^*},\, \right )^T,
\end{gather}
$\tilde J:T^*(W^{2N|2N})\to T(W^{2N|2N})$, corresponding to the
odd symplectic form $\omega^{(2)}=\sum\limits_{i=1}^N(\tilde
f_i\wedge \tilde f_i^*- \tilde \Phi_i\wedge \tilde \Phi_i^*)$, at
a point $(\tilde f_i,\tilde \Phi_i^*,\tilde f_i^*,\tilde
\Phi)^T\in W^{2N|2N}$. It should be noted that the Poisson
structure~\eqref{eq13} generates equation~\eqref{eq7} for any
Casimir functional $\gamma\in I({\cal G}^*)$.

Thus, on the extended phase space ${\cal G}^*\oplus W^{2N|2N}$ one
can obtain a Poisson structure as the tensor product $\tilde{\cal
L} :=\tilde\Theta\otimes \tilde J$ of \eqref{eq13} and
\eqref{eq14}.

Consider the following B\"acklund transformation:
\begin{gather}\label{eq15}
(\tilde l, \tilde f_i, \tilde \Phi_i^*, \tilde f_i^*,
\tilde\Phi_i)^T: \stackrel{B}{\mapsto} (l(\tilde l, \tilde f_i,
\tilde \Phi_i^*, \tilde f_i^*,\tilde \Phi_i), f_i =\tilde f_i,
\Phi_i^*=\tilde \Phi_i^*, f_i^*=\tilde f_i^*, \Phi_i= \tilde
\Phi_i)^T,
\end{gather}
generating on ${\cal G}^*\oplus W^{2N|2N}$ a Poisson structure
${\cal L}$ with respect to variables
$(l,f_i,\Phi_i^*,f_i^*,\Phi_i)$, $i=\overline{1,N}$, of the
coupled evolution equations \eqref{eq7}, \eqref{eq11} and
\eqref{eq12}. The main condition for the mapping \eqref{eq15} is
coincidence of the dynamical system
\begin{gather}\label{eq16}
(d l/dt,\,df_i/dt,\, d\Phi_i^*/dt,\, \,df_i^*/dt,\,
d\Phi_i/dt)^T:=-{\cal L} \nabla\gamma_n ( l, f_i, \Phi_i^*, f_i^*,
\Phi_i)
\end{gather}
with equations \eqref{eq7}, \eqref{eq11} and \eqref{eq12} in the
case of $\gamma_n\in I ({\cal G}^*) $, $n\in \ZZ_+$, i.e.\ when
the functional $\gamma_n$ is taken to be not dependent of
variables $(f_i,\Phi_i^*,f_i^*,\Phi_i)^T\in W^{2N|2N}$. To satisfy
that condition, one should find a variation of some Casimir
functional $\gamma_n\in I ({\cal G}^*) $, $n\in \ZZ_+$, at
$\delta\tilde l=0$, taking into account the evolutions
\eqref{eq11}, \eqref{eq12} and the B\"acklund transformation
\eqref{eq15}:
\begin{gather}
\left . \delta\gamma_n (\tilde l, \tilde f_i, \tilde \Phi_i^*,
\tilde f_i^*,\tilde \Phi_i)\right |_{\delta\tilde l=0}   \nonumber\\
\qquad {} =\sum_{i=1}^N \left ( \langle\delta\tilde f_i
,\frac{\delta\gamma_n}{\delta\tilde f_i}\rangle+
\langle\delta\tilde \Phi_i^*, \frac{\delta\gamma_n}{\delta\tilde
\Phi_i^*}\rangle + \langle\delta\tilde f_i^*
,\frac{\delta\gamma_n}{\delta\tilde f_i^*}\rangle
+\langle\delta\tilde \Phi_i ,\frac{\delta\gamma_n}{\delta\tilde
\Phi_i}\rangle\right )
\nonumber\\
\qquad {} =\sum_{i=1}^N \left ( \langle \delta\tilde f_i
,-\frac{d\tilde f_i^*}{dt_n}\rangle+ \langle \delta\tilde \Phi_i^*
,\frac{d\tilde \Phi_i}{dt_n}\rangle+ \langle\delta\tilde f_i^*,
d\tilde f_i /dt_n\rangle + \langle\delta\tilde \Phi_i
,-\frac{d\tilde \Phi_i^*}{dt_n}\rangle \right )
\nonumber \\
\qquad {} =\sum_{i=1}^N \biggl ( \langle \delta f_i,
(\nabla\gamma_n (l))_+^* f_i^*\rangle+ \langle\delta \Phi_i^*,
(\nabla\gamma_n (l))_+ \Phi_i\rangle+ \langle\delta f_i^*,
(\nabla\gamma_n (l))_+ f_i\rangle
\nonumber \\
\qquad \phantom{=}{}+ \langle\delta \Phi_i, (\nabla\gamma_n
(l))_+^* \Phi_i^*\rangle \biggr ) = \sum_{i=1}^N \biggl (
\langle(\nabla\gamma_n (l))_+ \delta f_i, f_i^*\rangle +\langle(\nabla\gamma_n (l))_+ f_i, \delta f_i^*\rangle\nonumber \\
\qquad \phantom{=}{} +\langle(\nabla\gamma_n (l))_+(\delta
\Phi_i), \Phi_i^*\rangle+ \langle(\nabla\gamma_n (l))_+ \Phi_i,
\delta \Phi_i^*\rangle
\biggr ) \nonumber \\
\qquad {} =\sum_{i=1}^N \biggl ( (\nabla\gamma_n (l),\delta (f_i
D_{\theta}^{-1} f_i^*))+ (\nabla\gamma_n (l), \delta (\Phi_i
D_{\theta}^{-1} \Phi_i^*))  \biggr )
\nonumber \\
\qquad {} =\biggl (\nabla\gamma_n (l),\delta \sum_{i=1}^N(f_i
D_{\theta}^{-1} f_i^*+ \Phi_i D_{\theta}^{-1} \Phi_i^*)\biggr ) :=
(\nabla\gamma_n (l),\delta l),\label{eq17}
\end{gather}
where $\gamma_n\in I ({\cal G}^*) $, $n\in \ZZ_+$, at the point $
l\in {\cal G}^*$ and the brackets $\langle \cdot,\cdot\rangle$
designate paring of the spaces $W^{1|0}$ and $W^{0|1}$. As a
result of the expression \eqref{eq17} one obtains the
relationships:
\[
\left . \delta l \right |_{\delta \tilde l=0} = \delta
\sum_{i=1}^N(f_i D_{\theta}^{-1} f_i^*+ \Phi_i D_{\theta}^{-1}
\Phi_i^*).
\]
Having assumed the linear dependence of $l$ from $\tilde l\in
{\cal G}^*$ one gets right away that
\begin{gather}\label{eq18}
l=\tilde l+\sum_{i=1}^N(f_i D_{\theta}^{-1} f_i^*+ \Phi_i
D_{\theta}^{-1} \Phi_i^* ).
\end{gather}
Thus, the B\"acklund transformation \eqref{eq15} can be written as
\begin{gather}\label{eq19}
(\tilde l, \tilde f_i, \tilde \Phi_i^*, \tilde f_i^*,
\tilde\Phi_i)^T \stackrel{B}{\mapsto} \biggl ( l=\tilde l
+\sum_{i=1}^N  (f_i D_{\theta}^{-1} f_i^*+ \Phi_i D_{\theta}^{-1}
\Phi_i^* ), f_i, \Phi_i^*, f_i^*, \Phi_i \biggr )^T.
\end{gather}
The expression \eqref{eq19} generalizes the result obtained in the
papers~\cite{hentosh10,hentosh11,hentosh12} for the Lie algebra of
integral-differential operators. The existence of the B\"acklund
transformation \eqref{eq19} makes it possible to formulate the
following theorem.
\begin{theorem}\label{theorem1}
The dynamical system on ${\cal G}^*\oplus W^{2N|2N}$, being
Hamiltonian with respect to the Poisson structure $\tilde {\cal L}
:T^*({\cal G}^*\oplus W^{2N|2N})\to T({\cal G}^*\oplus
W^{2N|2N})$, in the form of the following evolution equations:
\begin{gather*}
\frac {d\tilde l}{dt_n}= \left [\left
(\frac{\delta\gamma_n}{\delta\tilde l}\right )_+, \tilde l \right
] - \left [\frac{\delta\gamma_n}{\delta\tilde l}, \tilde l \right
]_+,
\\
\frac {d\tilde f_i}{dt_n}=\frac{\delta\gamma_n}{\delta\tilde
f_i^*}, \quad \frac {d\tilde
\Phi_i^*}{dt_n}=-\frac{\delta\gamma_n}{\delta\tilde \Phi_i}, \quad
\frac {d\tilde f_i^*}{dt_n}=-\frac{\delta\gamma_n}{\delta\tilde
f_i}, \quad \frac {d\tilde
\Phi_i}{dt_n}=\frac{\delta\gamma_n}{\delta\tilde \Phi_i^*},
\end{gather*}
where $i=\overline{1,N}$ and $\gamma_n\in I({\cal G}^*)$, $n\in
\ZZ_+$, is a Casimir functional at the point $l\in {\cal G}^*$,
connected with $\tilde l\in {\cal G}^*$ by \eqref{eq18}, is
equivalent to the system \eqref{eq9}, \eqref{eq13} and
\eqref{eq14} via the B\"acklund transformation \eqref{eq19}.
\end{theorem}

By means of simple calculations via the formula:
\[
{\cal L}  = B^{'} \tilde {\cal L}  B^{'*},
\]
where $B^{'}: T({\cal G}^*\oplus W^{2N|2N})\to T({\cal G}^*\oplus
W^{2N|2N})$ is a Frechet derivative of \eqref{eq19}, one brings
about the following form of the Poisson structure ${\cal L} $ on
${\cal G}^*\oplus W^{2N|2N} \ni ( l,f_i,\Phi_i^*,f_i^*,\Phi_i)^T$:
\begin{gather}\label{eq20}
\nabla\gamma (l, f_i,\Phi_i^*, f_i^*,\Phi_i)\stackrel{{\cal L}
}{\to} \left( \begin{array}{c} \left [\tilde l ,\left
(\D\frac{\delta\gamma}{\delta\tilde l}\right )_+\right ] - \left
[\tilde l ,\D\frac{\delta\gamma}{\delta\tilde l}\right ]_+
+\D\sum_{i=1}^N\biggl ( f_i D_{\theta}^{-1}\D\frac{\delta\gamma}{\delta f_i}-\\
-\D\frac{\delta\gamma}{\delta f_i^*} D_{\theta}^{-1}f_i^*+ \Phi_i
D_{\theta}^{-1}\D\frac{\delta\gamma}{\delta \Phi_i}-
\D\frac{\delta\gamma}{\delta \Phi_i^*} D_{\theta}^{-1}\Phi_i^*
\biggr ) \\
-\D\frac{\delta\gamma}{\delta f_i^*}-\left (\frac{\delta\gamma}{\delta l}\right )_+ f_i \\
\D\frac{\delta\gamma}{\delta \Phi_i}+\left (\frac{\delta\gamma}{\delta l}\right )_+^* \Phi_i^* \\
\D\frac{\delta\gamma}{\delta f_i^*}+\left (\frac{\delta\gamma}{\delta l}\right )_+^* f_i^* \\
-\D\frac{\delta\gamma}{\delta \Phi_i^*}-\left
(\frac{\delta\gamma}{\delta l}\right )_+ \Phi_i
\end{array}\right ),
\end{gather}
where $\gamma\in D({\cal G}^*\oplus W^{2N|2N})$ is an arbitrary
smooth functional and $i=\overline{1,N}$, that makes it possible
to formulate the theorem.

\begin{theorem}\label{theorem2}
For every $n\in \ZZ_+$ the coupled dynamical system \eqref{eq7},
\eqref{eq11} and \eqref{eq12} is Hamiltonian with respect to the
Poisson structure ${\cal L} $ in the form \eqref{eq20} and the
functional $\gamma_n\in I ( {\cal G}^*) $.
\end{theorem}

Using the expression \eqref{eq18}
 one can construct the hierarchy of Hamiltonian evolution equations, describing commutative flows,
 generated by involutive with respect to the Lie--Poisson bracket~\eqref{eq3}
 Casimir invariants $\gamma_n\in I({\cal G}^*)$, $n\in \ZZ_+$,
on the extended space ${\cal G}^*\oplus W^{2N|2N}$ at a~fixed
element $\tilde l\in {\cal G}^*$. For every $n\in \ZZ_+$ the
equation of such a type is equivalent to the system \eqref{eq7},
\eqref{eq11} and~\eqref{eq12}.

\section{Hierarchies of additional symmetries}

The evolution type hierarchy \eqref{eq7}, \eqref{eq11} and
\eqref{eq12} possesses another set of invariants, which includes
all natural powers of the eigenvalues $\lambda_i$,
$i=\overline{1,N}$. They can be considered as smooth by Frechet
functionals on the extended space ${\cal G}^*\oplus W^{2N|2N}$ due
to the representation:
\begin{gather}\label{eq21}
\lambda_k^s=\langle l^s f_k,f_k^*\rangle+\langle l^s
\Phi_k,\Phi_k^*\rangle,
\end{gather}
where $s\in \NN$, taking place for all $k=\overline{1,N}$ under
the normalizing condition:
\[
\langle f_k,f_k^*\rangle+\langle \Phi_k,\Phi_k^*\rangle=1.
\]
In the case of
\begin{gather}\label{eq22}
l:= l_+ + \sum_{i=1}^N(f_i D_{\theta}^{-1} f_i^*+ \Phi_i
D_{\theta}^{-1} \Phi_i^* )
\end{gather}
the formula \eqref{eq21} leads to the following variation of the
functionals $\lambda_k^s\in {\cal D}({\cal G}^*\oplus W^{2N|2N})$,
$k=\overline{1,N}$:
\begin{gather*}
\delta \lambda_k^s=\langle (\delta l^s)
f_k,f_k^*\rangle+\langle(\delta l^s) \Phi_k, \Phi_k^*\rangle
\\
\phantom{\delta \lambda_k^s=} {} +\langle l^s (\delta
f_k),f_k^*\rangle+ \langle l^s f_k,\delta f_k^*\rangle+
\langle l^s (\delta \Phi_k), \Phi_k^*\rangle + \langle  l^s \Phi_k,\delta \Phi_k^*\rangle \\
\phantom{\delta \lambda_k^s} {} =(\delta l_+, M_k^s)+ \sum_{i=1}^N
\bigl ( \langle \delta f_i, (-M_k^s+\delta_k^i l^s)^*f_i^*\rangle
+\langle\delta f_i^*, (-M_k^n +\delta_k^i l^s) f_i\rangle  \\
\phantom{\delta \lambda_k^s=} {} +\langle \delta \Phi_i,
(-M_k^s+\delta_k^i l^s)^*\Phi_i^*\rangle + \langle\delta \Phi_i^*,
(-M_k^n +\delta_k^i l^s) \Phi_i\rangle \bigr ),
\end{gather*}
where $\delta_k^i$ is a Kronecker symbol and the operator $M_k^s$,
$s\in \NN$, is determined as
\[
M_k^s:= \sum_{p=0}^{s-1}\bigl ( ( l^pf_k)D_{\theta}^{-1}
(l^{*s-1-p} f_k^*)+ ( l^p \Phi_k)D_{\theta}^{-1} (l^{*s-1-p}
\Phi_k^*)\bigr )=\lambda_k^{s-1} M_k^1.
\]
Thus, one obtains the exact forms of gradients for the functionals
$\lambda_k^s\in {\cal D}(\hat {\cal G}^*\oplus W^{2N|2N})$,
$k=\overline{1,N}$:
\begin{gather}\label{eq23}
\nabla \lambda_k^s(l_+,f_i,\Phi_i^*,f_i^*, \Phi_i)= \left (
\begin{array}{c}
M_k^s \\
(-M_k^s+\delta_k^i l^s)^*f_i^* \\
(-M_k^n +\delta_k^i l^s) \Phi_i \\
(-M_k^n +\delta_k^i l^s) f_i \\
(-M_k^s+\delta_k^i l^s)^*\Phi_i^*
\end{array} \right ),
\end{gather}
where $i=\overline{1,N}$. By means of the expression \eqref{eq23}
the tensor product $\tilde {\cal L} $ of Poisson
structures~\eqref{eq13} and \eqref{eq14} generates the hierarchy
of coupled evolution equations on ${\cal G}^*\oplus W^{2N|2N}$:
\begin{gather}
d l_+/d\tau _{s,k}=-[M_k^s,\hat l_+]_+,\label{eq24}
\\
df_i/d\tau _{s,k} = (-M_k^s+\delta_k^i l^s) f_i, \qquad
df^*_i/d\tau _{s,k} = (M_k^s-\delta_k^i  l^s )^*
f_i^*,\label{eq25}
\\
d \Phi_i/d\tau _{s,k} = (-M_k^s+\delta_k^i l^s) \Phi_i, \qquad d
\Phi^*_i/d\tau _{s,k} = (M_k^s-\delta_k^i  l^s )^*
\Phi_i^*,\label{eq26}
\end{gather}
where $i=\overline{1,N}$, for every $k=\overline{1,N}$. Because of
the B\"acklund transformation \eqref{eq19} the
equation~\eqref{eq24} is equivalent to the commutator
relationship:
\begin{gather}\label{eq27}
d l/d\tau _{s,k}=-[M_k^s, l ]=-\lambda_k^{s-1}[M_k^1,
l]=\lambda_k^{s-1}d l/d\tau _{1,k},
\end{gather}
and the following theorem takes place:
\begin{theorem}\label{theorem3}
For every $k=\overline{1,N}$ and $s\in \NN$ the coupled dynamical
system \eqref{eq24}, \eqref{eq25} and \eqref{eq26} is Hamiltonian
one with respect to the Poisson structure ${\cal L} $ in the form
\eqref{eq20} and the functional $\lambda_k^n\in {\cal D}( {\cal
G}^*\oplus W^{2N|2N})$.
\end{theorem}

The coupled dynamical systems \eqref{eq24}, \eqref{eq25} and
\eqref{eq26} represent flows on ${\cal G}^*\oplus W^{2N|2N}$,
commuting one with each other.
\begin{theorem}\label{theorem4}
For $k=\overline{1,N}$ the coupled evolution equations
\eqref{eq24}, \eqref{eq25} and \eqref{eq26} form a set of
additional symmetry hierarchies for the coupled dynamical system
\eqref{eq7}, \eqref{eq11} and \eqref{eq12}.
\end{theorem}

\begin{proof}
To prove the theorem it is sufficient to show that
\begin{gather}\label{eq28}
[d/dt_n, d/d\tau_{1,k}]=0, \qquad [d/d \tau_{1,k},
d/d\tau_{1,q}]=0,
\end{gather}
where $k,q=\overline{1,N}$ and $n\in \NN$. The first equality in
the formula \eqref{eq28} follows from the identities:
\[
d (\nabla \gamma_n(l))_+/d \tau_{1,k}=[(\nabla \gamma_n(l))_+,
M_1^1]_+, \qquad d M_1^1/dt_n =[(\nabla \gamma_n(\hat l))_+,
M_1^1]_-,
\]
the second one being a consequence of the relationship:
\begin{equation*}
d M_k^1/d\tau_{1,q}-d M_q^1/d\tau_{1,k}=[M_k^1,M_q^1].
\tag*{\qed}
\end{equation*}
\renewcommand{\qed}{}
\end{proof}

When $N\ge 2$, a new class of nontrivial Hamiltonian flows
$d/dT_{n,K} :=d/dt_n+\sum\limits_{k=1}^{K}d/d\tau_{n,k}$, $n\in
\NN$, $K=\overline{1,N-1}$, in a Lax form on ${\cal G}^*\oplus
W^{2N|2N}$ can be constructed by use of the set of additional
symmetry hierarchies for the Lie algebra of
super-integro-differential operators. Acting on the functions
$f_i$, $f^*_i$, $\Phi_i$, $\Phi^*_i$, $i=\overline{1,N}$, these
flows generate ($(1+K)|$1+1)-dimensional supersymmetric nonlinear
dynamical systems.

For the first time the additional symmetries in the case of $N=2$
were applied by E.~Nissimov and S.~Pacheva~\cite{hentosh12a} to
obtain a Lax integrable supersymmetric analog of the
$(2+1)$-dimensional Davey--Stewartson system. If
\[
l:=\partial+f_1D_{\theta} ^{-1} f_1^*+f_2D_{\theta} ^{-1}
f_2^*+\Phi_1D_{\theta} ^{-1} \Phi_1^*+\Phi_2D_{\theta} ^{-1}
\Phi_2^* \in {\cal G}^*,
\]
where $(f_1,f_2,\Phi^*_1,\Phi^*_2, f^*_1,f^*_2, \Phi_1, \Phi_2)^T
\in W^{4|4}$, the flows $\partial / \partial\tau :=d/d\tau_{1,1}$
and $d/dT :=d/dT_{2,1} =d/dt_2+d/d\tau_{2,1}$ on ${\cal G}^*\oplus
W^{4|4}$, acting on the functions $f_i$, $f^*_i$, $\Phi_i$,
$\Phi^*_i$, $i=1,2$, by the following way:
\begin{gather}
f_{1,\tau}=f_{1,x}+u_1f_2-\alpha_1\Phi_2, \qquad
f_{2,\tau}=-\bar u_1f_1+\bar \alpha_2\Phi_1, \nonumber \\
f^*_{1,\tau}=f^*_{1,x}+\bar u_1f^*_2-\bar\alpha_1\Phi^*_2, \qquad
f^*_{2,\tau}=- u_1f^*_1- \alpha_2\Phi^*_1, \nonumber \\
\Phi_{1,\tau}=\Phi_{1,x}-\alpha_2f_2+u_2\Phi_2, \qquad
\Phi_{2,\tau}=-\bar \alpha_1f_1-\bar u_2\Phi_1, \nonumber \\
\Phi^*_{1,\tau}=\Phi^*_{1,x}-\bar \alpha_2f^*_2-\bar u_2\Phi^*_2,
\qquad \Phi^*_{2,\tau}= \alpha_1f^*_1- u_2\Phi^*_1,\label{eq29}
\end{gather}
and
\begin{gather}
f_{1,T}=f_{1,xx}+f_{1,\tau\tau} + w_1D_{\theta}f_1+w_0f_1+2v_{1,\tau}f_1-2\beta _{\tau} \Phi_1, \nonumber \\
f_{2,T}=f_{2,xx}+ w_1D_{\theta}f_2+w_0f_2- \bar u_1 f_{1,\tau}
+\bar \alpha_2 \Phi_{1,\tau}+\bar u_{1,\tau} f_1 -
\bar \alpha_{2,\tau} \Phi_{1,\tau}, \nonumber \\
f^*_{1,T}=-f^*_{1,xx}-f^*_{1,\tau\tau}-D_{\theta}(w_1f^*_1)-w_0f^*_1-
2v_{1,\tau}f^*_1+2\bar \beta _{\tau} \Phi^*_1, \nonumber \\
f^*_{2,T}=-f^*_{2,xx}-D_{\theta} (w_1 f^*_2)-w_0f^*_2+ u_1
f^*_{1,\tau} +\alpha_2 \Phi^*_{1,\tau}- u_{1,\tau} f^*_1 +
\alpha_{2,\tau} \Phi^*_{1,\tau} , \nonumber \\
\Phi_{1,T}=\Phi_{1,xx}+\Phi_{1,\tau\tau} +
w_1D_{\theta}\Phi_1+w_0\Phi_1+
2f_1\bar\beta _{\tau}+2v_{2,\tau}\Phi_1, \nonumber \\
\Phi_{2,T}=\Phi_{2,xx}+ w_1D_{\theta}\Phi_2+w_0\Phi_2 - \bar
\alpha_1 f_{1,\tau} -\bar u_2 \Phi_{1,\tau}+\bar \alpha_{1,\tau}
f_1 +
\bar u_{2,\tau} \Phi_{1,\tau} , \nonumber \\
\Phi^*_{1,T}=-\Phi^*_{1,xx}-\Phi_{1,\tau\tau} - D_{\theta}(w_1\Phi^*_1)-w_0\Phi^*_1-2f^*_1 \beta _{\tau}-2v_{2,\tau}\Phi^*_1, \nonumber \\
\Phi^*_{2,T}=-\Phi^*_{2,xx}-D_{\theta} (w_1\Phi^*_2)-w_0\Phi^*_2-
\alpha_1 f^*_{1,\tau} + u_2 \Phi^*_{1,\tau}+ \alpha_{1,\tau} f^*_1
-
u_{2,\tau} \Phi^*_{1,\tau}, \nonumber \\
D_{\theta}u_1=f_1f_2^*,\qquad  D_{\theta}u_2=\Phi_1\Phi^*_2,
\qquad
D_{\theta}\bar u_1=f^*_1f_2,\qquad  D_{\theta}\bar u_2=\Phi_1^*\Phi_2, \nonumber \\
D_{\theta}v_1=f_1f_1^*, \qquad D_{\theta}v_2=\Phi_1\Phi^*_1, \qquad
D_{\theta}\alpha_1=f_1\Phi_2^*,\qquad  D_{\theta}\alpha_2=\Phi_1
f^*_2, \nonumber \\
D_{\theta}\bar \alpha_1=f^*_1\Phi_2,\qquad  D_{\theta}\bar \alpha_2=\Phi^*_1 f_2, \qquad
D_{\theta}\beta=f_1\Phi_1^*, \qquad D_{\theta}\bar
\beta=f^*_1\Phi_1,\label{eq30}
\end{gather}
where $(\nabla\gamma_2 ( l))_+:=\partial^2+w_1D_{\theta}+w_0$,
represent $(2|1+1)$-dimensional supersymmetric nonlinear dynamical
system. The system \eqref{eq29} and \eqref{eq30} possesses an
infinite sequence of local conservation laws, which can be found
by the formula~\eqref{eq6}, and a Lax representation, given by the
spectral problem~\eqref{eq8} and the evolution equations:
\begin{gather}
f_{\tau}=-M_1^1f, \label{eq31}\\
f_T=((\nabla\gamma_2 (l))_+-M_1^2)f,\label{eq32}
\end{gather}
for an arbitrary eigenfunction $f\in W^{1|0}$ or $f\in W^{0|1}$.
The relationships~\label{eq31} and~\eqref{eq32} lead to additional
nonlinear constraints such as
\begin{gather}
w_{0,\tau}=2w_1(f_1f_1^*-\Phi_1 \Phi_1^*)+2(f_1 (D_{\theta}f_1^*)+\Phi_1 (D_{\theta}\Phi_1^*))_x , \nonumber \\
w_{1,\tau}=-2(f_1f_1^*)_x+2(\Phi_1 \Phi_1^*)_x.\label{eq33}
\end{gather}
When $f_1:=\psi$, $f^*_1:=\theta \psi^*$, $f_2=f^*_2=0$ and
$\Phi_1=\Phi^*_1=\Phi_2=\Phi^*_2=0$, the equations \label{eq30}
and \label{eq33} are reduced to the Lax integrable
$(2+1)$-dimensional Davey--Stewartson
system~\cite{hentosh5,hentosh16,hentosh17}:
\begin{gather*}
\psi_{1,T}=\psi_{1,xx}+\psi_{1,\tau\tau} + 2(S-2\psi\psi^*)\psi , \\
\psi^*_{1,T}=-\psi^*_{1,xx}-\psi^*_{1,\tau\tau} - 2(S-2\psi\psi^*)\psi^* , \\
S_{x\tau}=(\partial/\partial x+\partial/\partial \tau)^2
\psi\psi^*,
\end{gather*}
where $2S:=w_0^0+2 v^0_{1,\tau}+4\psi\psi^*$, $w_0:=w_0^0$,
$v_{1,\tau}:=v^0_{1,\tau}$ and $\psi,\psi^* \in L_\infty
(\SS^1;\CC)$.

The Lax representation~\eqref{eq10}, \eqref{eq31} and \eqref{eq32}
for the $(2|1+1)$-dimensional supersymmetric nonlinear dynamical
Davey--Stewartson system~\eqref{eq29}, \eqref{eq30} and
\eqref{eq33} has equivalent matrix form:
\begin{gather*}
D_{\theta} F= \left (\begin{array}{cccccc}
0 & 0 & 0 & 0 & 0 & 1 \\
f_1^* & 0 & 0 & 0 & 0 & 0 \\
f_2^* & 0 & 0 & 0 & 0 & 0 \\
\Phi_1^* & 0 & 0 & 0 & 0 & 0 \\
\Phi_2^* & 0 & 0 & 0 & 0 & 0 \\
\lambda & -f_1 & -f_2 & -\Phi_1 & -\Phi_2 & 0
\end{array}\right )F,
\\
\D \frac {dF}{d\tau}= \left (\begin{array}{cccccc}
0 & -f_1 & 0 & -\Phi_1 & 0 & 0 \\
D_{\theta}f^*_1 & -\lambda & \bar u_1 & 0 & \bar \alpha_1 & -f^*_1 \\
0 & -u_1 & 0 & \alpha_2 & 0 & 0 \\
D_{\theta}\Phi^*_1 & 0 & \bar \alpha_2 & -\lambda & \bar u_2 & \Phi^*_1 \\
0 & -\alpha_1 & 0 & -u_2 & 0 & 0 \\
\Phi_1\Phi^*_1-f_1f^*_1 & -D_{\theta} f_1 & 0 & -D_{\theta} \Phi_1
& 0 & 0
\end{array}\right )F,
\\
\D \frac {dF}{dT}=CF,
\end{gather*}
where $F=(F^0:=f,F^2,F^4, F^1,F^3,F^5)^\tau\in W^{3|3}$,
$C:=(C_{mn})\in {\rm gl}(3|3)$, $m,n=\overline{1,6}$, and
\begin{gather*}
C_{11}=\lambda^2+\frac 12 w_0 + f_1 D_{\theta} f^*_1 + \Phi_1
D_{\theta} \Phi^*_1, \qquad
C_{12}=-(2\lambda f_1 + f_{1,x}+f_{1,\tau}),  \\
C_{13}=-(\lambda f_2 + f_{2,x})+\bar u_1 f_1-\bar \alpha_2 \Phi_1,
\qquad
C_{14}=-(2\lambda \Phi_1 + \Phi_{1,x}+\Phi_{1,\tau}),  \\
C_{15}=-(\lambda \Phi_2 + \Phi_{2,x})-\bar u_2 \Phi_1+\bar
\alpha_1 f_1, \qquad
C_{16}=\frac 12 w_1 - f_1 f^*_1 + \Phi_1 \Phi^*_1, \\
C_{21}=-w_1 f^*_1+2D_{\theta}(- f^*_{1,x}+\lambda f^*_1)-
\bar u_1 D_{\theta}f^*_2 - \bar \alpha_1 D_{\theta}\Phi^*_2 , \\
C_{22}=-\lambda ^2-2D_{\theta}(f_1 f_1^*)-u_1\bar u_1+
\alpha_1\bar \alpha_1 ,  \qquad
C_{23}=-D_{\theta}(f_2 f_1^*)  +\lambda\bar u_1-\bar u_{1,\tau}, \\
C_{24}=-2D_{\theta}(\Phi_1 f_1^*)+\bar u_1\alpha_2-u_2 \bar
\alpha_1, \qquad
C_{25}=-D_{\theta}(\Phi_2 f_1^*)+\lambda \bar \alpha_1-\bar \alpha_{1,\tau}, \\
C_{26}=2(-\lambda f^*_1 + f^*_{1,x})+\bar u_1 f^*_2 -\bar
\alpha_1\Phi^*_2,
\\
C_{31}=-\frac 12 w_1 f^*_2+D_{\theta} (- f^*_{2,x}+\lambda f^*_2)+
u_1 D_{\theta}f^*_1 + \alpha_2 D_{\theta}\Phi^*_1 , \\
C_{32}=-D_{\theta}(f_1 f_2^*)-\lambda u_1-u_{1,\tau} ,  \qquad
C_{33}=-D_{\theta}(f_2 f_2^*)  +u_1\bar u_1-\alpha_2\bar \alpha_2, \\
C_{34}=-D_{\theta}(\Phi_1 f_2^*)+\lambda \alpha_2+\alpha_{2,\tau}
, \qquad
C_{35}=-D_{\theta}(\Phi_2 f_2^*)+u_1 \bar \alpha_1 -\bar u_2\alpha_2 , \\
C_{36}=(-\lambda f^*_2 + f^*_{2,x})-u_1 f^*_1+\alpha_2\Phi^*_1,
\\
C_{41}=-w_1 \Phi^*_1+2 D_{\theta}(-\Phi^*_{1,x}+\lambda \Phi^*_1)
+\bar u_2 D_{\theta}\Phi^*_2 - \bar \alpha_2 D_{\theta}f^*_2 , \\
C_{42}=-2D_{\theta}(f_1 \Phi_1^*)-u_1\bar \alpha_2- \bar u_2
\alpha_1 ,  \qquad
C_{43}=-D_{\theta}(f_2 \Phi_1^*)  +\lambda\bar \alpha_2-\bar \alpha_{2,\tau},\\
C_{44}=-\lambda^2-2D_{\theta}(\Phi_1 \Phi_1^*)-u_2\bar
u_2-\alpha_2\bar \alpha_2, \qquad
C_{45}=-D_{\theta}(\Phi_2 \Phi_1^*)+\lambda \bar u_2-\bar u_{2,\tau},\\
C_{46}=2(-\lambda \Phi^*_1 + \Phi^*_{1,x})+\bar u_2 \Phi^*_2 +\bar
\alpha_2 f^*_2,
\\
C_{51}=-\frac 12 w_1 \Phi^*_2+ D_{\theta}(-\Phi^*_{2,x}+\lambda
\Phi^*_2)
+\alpha_1 D_{\theta}f^*_1 + u_2 D_{\theta}\Phi^*_1 , \\
C_{52}=-D_{\theta}(f_1 \Phi_2^*)-\lambda \alpha_1-\alpha_{1,\tau}
,  \qquad
C_{53}=-D_{\theta}(f_2 \Phi_2^*)  +\alpha_1\bar u_1-\bar\alpha_2 u_2, \\
C_{54}=-D_{\theta}(\Phi_1 \Phi_2^*)-\lambda u_2-u_{2,\tau} ,
\qquad
C_{55}=-D_{\theta}(\Phi_2 \Phi_2^*)+\alpha_1 \bar \alpha_1 +u_2\bar u_2 , \\
C_{56}=(-\lambda \Phi^*_2 + \Phi^*_{2,x})-\alpha_1 f^*_1+u_2\Phi^*_1, \\
C_{61}=\frac 12 D_{\theta}w_0 + (D_{\theta}f_1) D_{\theta} f^*_1 +
(D_{\theta}\Phi_1) D_{\theta} \Phi^*_1 \\
\phantom{C_{61}=}{}-(f_{1,\tau}f^*_1+f_{2,x}f^*_2- \Phi_{1,\tau}
\Phi^*_1-\Phi_{2,x} \Phi^*_2) +\bar u_1 f_1f^*_2 -\bar \alpha_1
f_1\Phi^*_2+\bar \alpha_2 \Phi_1f^*_2+\bar u_2 \Phi_1 \Phi^*_2,
\\
C_{62}=-D_{\theta}(2\lambda f_1 + f_{1,x}+f_{1,\tau})+\frac 12 w_1f_1+f_1(-f_1f^*_1+\Phi_1 \Phi^*_1),  \\
C_{63}=-D_{\theta}(\lambda f_2 + f_{2,x})+\frac 12 w_1f_2+ \bar u_1(D_{\theta} f_1)+\bar \alpha_2(D_{\theta} \Phi_1) \\
C_{64}=-D_{\theta}(2\lambda \Phi_1 + \Phi_{1,x}+\Phi_{1,\tau})+\frac 12 w_1\Phi_1-f_1f^*_1\Phi_1,  \\
C_{65}=-D_{\theta}(\lambda \Phi_2 + \Phi_{2,x})+\frac 12 w_1\Phi_2
-\bar u_2 D_{\theta}\Phi_1-\bar \alpha_1 D_{\theta}f_1, \\
C_{66}=\lambda^2+\frac 12 w_0+\frac 12 D_{\theta}w_1-(D_{\theta}
f_1)f^*_1+(D_{\theta}\Phi_1) \Phi^*_1 .
\end{gather*}
In fact, one has found a triple matrix linearization for a
$(2|1+1)$-dimensional dynamical system, that is important for the
standard method of inverse scattering
transformation~\cite{hentosh3} as well as for the reduction
procedure~\cite{hentosh18,hentosh19} upon invariant subspaces of
associated spectral problem eigenvalues.

The method of additional symmetries is effective for constructing
a wide class of $(2|1+1)$-dimensional supersymmetric nonlinear
dynamical systems with a triple matrix linearization.

\vspace{-1mm}

\section{Conclusion}

By now several regular Lie-algebraic approaches existed to
constructing Lax integrable \mbox{$(2+1)$}-dimensional nonlinear
dynamical systems on functional manifolds, which were presented
in~\cite{hentosh12,hentosh20,hentosh21,hentosh22}. In this paper a
new Lie-algebraic method is devised for introducing one more
commuting variable into $(1|1+1)$-dimensional dynamical systems
with preserving their integrability by Lax. It involves use of
additional symmetries~\cite{hentosh12a} for a Hamiltonian flow
hierarchy on extended dual space to some operator Lie algebra.

Any integrable $(2|1+1)$-dimensional supersymmetric nonlinear
dynamical system obtained by means of the method possesses an
infinite sequence of local conservation laws and a triple mat\-rix
linearization of a Lax type. These properties make it possible to
apply the standard inverse scattering
transformation~\cite{hentosh3} and the reduction
procedure~\cite{hentosh18,hentosh19} upon invariant subspaces.

If $N>2$ in the representation~\eqref{eq22}, the hierarchies of
additional symmetries can be used for constructing Lax integrable
($(1+K)|1+1)$-dimensional supersymmetric systems, where
$K=\overline{1,N-1}$.

Analyzing the structure of the B\"acklund type transformation
\eqref{eq19} as a key point of the method, one can observe that it
strongly depends on an $ad$-invariant scalar product chosen for an
operator Lie algebra ${\cal G}$ and a Lie algebra decomposition
like (2). Since there are other possibilities of choosing
$ad$-invariant scalar products on ${\cal G}$ and such
decompositions, they give rise naturally to other B\"acklund
transformations.

\subsection*{Acknowledgements}
The author thanks Professor A.K.~Prykarpatsky for useful
discussions and Organizers of Sixth International Conference
``Symmetry in Nonlinear Mathematical Physics'' (2005, Kyiv) for
invitation to take part in the conference. The present paper is
the written version of the talk delivered by the author at this
conference.

\LastPageEnding

\end{document}